\documentclass[aps,chaos,twocolumn,groupedaddress,showpacs,showkeys]{revtex4}

\usepackage{graphicx}
\usepackage{dcolumn}
\usepackage{bm}
\usepackage{amssymb}

\begin{document}

\title{Chaotic synchronization of coupled electron-wave systems with backward waves}

\author{Alexander~E.~Hramov}
\email{aeh@cas.ssu.runnet.ru}
\author{Alexey~A.~Koronovskii}
\email{alkor@cas.ssu.runnet.ru}
\author{Pavel~V.~Popov}
\author{Irene~S.~Rempen}
\affiliation{Faculty of Nonlinear Processes, Saratov State
University, Astrakhanskaya, 83, Saratov, 410012, Russia}

\date{\today}

\begin{abstract}
The chaotic synchronization of two electron-wave media with
interacting backward waves and cubic phase nonlinearity is
investigated in the paper. To detect the chaotic synchronization
regime we use a new approach, the so-called time scale
synchronization [Chaos, \textbf{14} (3) 603--610 (2004)]. This
approach is based on the consideration of the infinite set of
chaotic signals' phases introduced by means of continuous wavelet
transform. The complex space-time dynamics of the active media and
mechanisms of the time scale synchronization appearance are
considered.
\end{abstract}



\maketitle

{\bf Synchronization of chaotic oscillators is one of the
fundamental phenomena of nonlinear dynamics. In this paper we
consider chaotic synchronization of two electron-wave media with
interacting backward waves and cubic phase nonlinearity. An
efficient approach to detecting of synchronous dynamics of
distributed systems is suggested based on the time scale
synchronization concept [Chaos, \textbf{14} (3) 603--610 (2004)].
The complex space-time dynamics of the active media and mechanisms
of the appearance of time scale synchronization are discussed.}

\section{Introduction}
\label{intro}

In last decade the investigation of chaotic synchronization has
become a task of great interest \cite{Rosenblum:1996_PhaseSynchro,
Pikovsky:2000_SynchroReview, Boccaletti:2002_ChaosSynchro,
Pikovsky:2002_SynhroBook}. It can be observed in a lot of chaotic
oscillators of different nature, including physical and biological
systems \cite{Pikovsky:2002_SynhroBook,Ticos:2000_PlasmaDischarge,
Anishchenko:2000_humanSynchro, Tass:2003_NeuroSynchro,
Prokhorov:2003_HumanSynchroPRE}. Chaotic synchronization  can be
used for secret signal
transmission~\cite{Murali:1993_SignalTransmission,
Chua:1997_Criptography, Yaowen:2002_ChaoticCommunication,
Garcia-Ojalvo:2001_ChaosCommunication}. It also seems very
important to study the appearance of chaotic synchronization in
distributed systems demonstrating space--time chaos
\cite{Kurths:1995_DistrSystems, Boccaletti:1999_ControllingChaos,
Bragard:2000, Ticos:2000_Plasma, Ahlers:2002, Valladares:2002,
Mendoza:2004}.
Four types of chaotic synchronization have been carried out for
chaotic systems with lumped parameters, namely phase
synchronization (PS), complete synchronization (CS), lag
synchronization (LS) and generalized synchronization (GS)
\cite{Pecora:1997_SynchroChaos, Pikovsky:2002_SynhroBook,
Pikovsky:2000_SynchroReview, Boccaletti:2002_ChaosSynchro}. On the
ranges of mentioned regimes the intermittence behaviour \cite
{Boccaletti:2000_IntermitLagSynchro} takes place.
To describe the phase synchronization the instantaneous phase
$\phi(t)$ of a chaotic continuous time series is usually
introduced~(see \cite{Rosenblum:1996_PhaseSynchro,
Pikovsky:2002_SynhroBook, Anishchenko:2002_SynchroEng,
Pikovsky:2000_SynchroReview}). The PS means the entrainment of
phases of chaotic signals, whereas their amplitudes remain chaotic
and uncorrelated. Simultaneously, if Fourier spectrum of a signal
$x(t)$ is complex (chaotic time series are characterized by
Fourier spectrum without the main frequency), it is not always
possible to describe the system dynamics with the help of one
instantaneous phase $\phi(t)$ (see
\cite{Anishchenko:2004_ChaosSynchro, Hramov:2004_Chaos} for
detail). In this case other methods can be applied to analyze the
non-phase coherent complex systems dynamics (e.g., via curvature
or via recurrence, see \cite{Osipov:2003_Synchro_Chaos,
Romano:2004_Phys.Lett.A} for detail).

In our works \cite {Koronovskii:2004_JETPLettersEngl,
Hramov:2004_Chaos} the new approach to the analysis of the chaotic
synchronization is described, based on examination of continuous
set of phases $\phi_s(t)$ of chaotic time series introduced with
the help of the continuous wavelet transform \cite
{Daubechies:1992_WVTBook, Kaiser:1994_Wvt, alkor:2003_WVTBookEng}.

Let us consider continuous wavelet transform of time series $x(t)$
\begin{equation}
W(s,t_0)=\int\limits_{-\infty}^{+\infty}x(t)\psi^*_{s,t_0}(t)\,dt,
\label{eq:WvtTrans}
\end{equation}
where $\psi_{s,t_0}(t)$ is the wavelet--function related to the
mother--wavelet $\psi_{0}(t)$ as
\begin{equation}
\psi_{s,t_0}(t)=\frac{1}{\sqrt{s}}\psi\left(\frac{t-t_0}{s}\right).
\label{eq:Wvt}
\end{equation}
In equation (\ref{eq:WvtTrans}) $x(t)$ is the time series
generated by a dynamical system. It may also contain the noise
component. The time scale $s$ corresponds to the width of the
wavelet function $\psi_{s,t_0}(t)$, and $t_0$ is the shift of
wavelet along the time axis, symbol ``$*$'' in~(\ref{eq:WvtTrans})
denotes complex conjugation. It should be noted that the time
scale $s$ is usually used instead of the frequency $f$ of Fourier
transform and can be considered as the quantity inversed to it.

The Morlet--wavelet~\cite{Grossman:1984_Morlet}
\begin{equation}
\psi_0(\eta)=\frac{1}{\sqrt[4]{\pi}}\exp(j\Omega_0\eta)\exp\left(\frac{-\eta^2}{2}\right)
\label{eq:Morlet}
\end{equation}
has been used as a mother--wavelet function. The choice of
parameter value $\Omega_0=2\pi$ provides the relation ${s=1/f}$
between the time scale $s$ of wavelet transform and frequency $f$
of Fourier transform.

The wavelet surface
\begin{equation}
W(s,t_0)=|W(s,t_0)|e^{j\phi_s(t_0)} \label{eq:WVT_Phase}
\end{equation}
describes the system's dynamics on every time scale $s$ at moment
$t_0$. The value of $|W(s,t_0)|$ indicates the presence and
intensity of the time scale $s$ mode in the time series $x(t)$ at
the moment of time $t_0$. It is also possible to consider the
quantity
\begin{equation}
\langle E(s)\rangle=\int|W(s,t_0)|^2\,dt_0, \label{eq:IntEnergy}
\end{equation}
which is the distribution of integral energy by time scales.

At the same time, the phase $\phi_s(t)=\arg\,W(s,t)$ is naturally
introduced for every time scale $s$. In other words, $\phi_s(t)$
is a continuous function of time $t$ and time scale $s$. It means
that it is possible to describe the behavior of each time scale
$s$ by means of its own phase $\phi_s(t)$. Let us consider the
dynamics of two coupled oscillators. If in the time series
$\mathbf{x}_{1,2}(t)$ generated by these systems there is a range
of time scales $s_1\leq s\leq s_2$ for which the phase locking
condition
\begin{equation}
|\phi_{s1}(t)-\phi_{s2}(t)|<\mathrm{const} \label{eq:PhaseLocking}
\end{equation}
is satisfied and the part of the wavelet spectrum energy in this
range is not equal to zero
\begin{equation}
E_{snhr}=\int\limits_{s_1}^{s_2}\langle E(s)\rangle\,ds>0,
\label{eq:SynchroEnergy}
\end{equation}
then we assert that \textit{time scale synchronization} (TSS)
between oscillators takes place \cite{Hramov:2004_Chaos}. The
condition (\ref{eq:PhaseLocking}) can be generalized to the case
of $m:n$ synchronization. In this case one has to use the
condition
\begin{equation}
|m\phi_{sn1}(t)-n\phi_{sm2}(t)|<\mathrm{const}
\label{eq:PhaseLockingMN}
\end{equation}
instead of (\ref{eq:PhaseLocking}). In this case the time scales
$s_{m1}$ in the first system and $s_{n2}$ in the second system
would satisfy the relation $s_{m2}/s_{n1}=m/n$.

In Reference \cite {Hramov:2004_Chaos} the new approach was
applied to the study of chaotic synchronization of coupled
R\"ossler systems and two Chua's systems. It was shown, that
different types of synchronous oscillations (PS, LS, ILS, CS and
GS) may be considered as particular cases of TSS. It seems rather
interesting to apply the concept of time scale synchronization to
the analysis of chaotic synchronization of the distributed systems
demonstrating space-time chaos.

In this paper we present the results of investigation of chaotic
synchronization in the coupled electron--wave systems with
backward electromagnetic wave and cubic nonlinearity, derived with
the help of the method of time scale synchronization.
The explored system is a simple model of microwave oscillator with
backward wave (BWO) finding wide practical application (see, for
example, \cite{Soohoo:1971_MicrowaveElectronicsBook,
book:1983_microwave, Granatstein:1987_Book,
Felch:1999_SmallAuthor}). For similar systems the complex chaotic
regimes of space--time oscillations are typical
\cite{Bezruchko:1979_BWOChaos, Ginzburg:1979_engl,
Levush:1992_BWO, Trubetskov:1996_CHAOS, Nusinovich:2001,
Dronov:2004_TWTChaos}.

The structure of the paper is as follows.
Section~\ref{Sct:BaseEq} contains the formalism describing
nonlinear interaction of electron beam and backward
electromagnetic wave.
Section~\ref{Sct:Synchro} describes the synchronization of two
unidirectional coupled electron-wave systems. We demonstrate the
efficiency of TSS method for the analysis of synchronization of
spatially extended systems and discuss the characteristics of
chaotic synchronization of such systems.
The quantitative measure of synchronization of spatially extended
systems is described in section~\ref{Sct:Measure}. In
section~\ref{Sct:SpaceDynamics} we consider complex space-time
dynamics of the unidirectional coupled electron-wave systems with
the help of the concept of time-scale synchronization.
The final conclusion is presented in section~\ref{Sct:Conclusion}.

\section{General formalism}
\label{Sct:BaseEq}

Let us consider a simple model of two bound electron--wave media
with backward wave and cubic phase nonlinearity.

One of the types of the active media of electron--wave nature is
based on resonance interaction of electron wave and backward
electromagnetic wave in the slow-wave structure
\cite{Soohoo:1971_MicrowaveElectronicsBook, book:1983_microwave,
Granatstein:1987_Book, TrueAeh:2003_microwave_electronics_1engl}.
The electron wave is really a perturbation of density and velocity
of charged particles of electron beam propagating with the
velocity $v_0$ along a slow-wave structure.

In the linear approximation such system conforms to a model of
interaction of two opposing linear waves with dispersion relation
\begin{equation}
 \omega_1=k_1, \qquad \omega_2=-k_2.
\end{equation}
In this case the linear model can be formulated as a
self-consistent system of two equations
\begin{equation}\label{ref_bwopp_01}
  \frac{\partial F}{\partial \tau}-\frac{\partial F}{\partial \xi}=-{A}I,
\end{equation}
\begin{equation}\label{ref_bwopp_02}
  \frac{\partial I}{\partial \tau}+\frac{\partial I}{\partial \xi}=-{A}F,
\end{equation}
where an absolute instability is observed. Here
$F=|F|\exp(j\varphi_F)$ and $I=|I|\exp(j\varphi_I)$ are the
normalized slowly variable amplitudes of interacting
electromagnetic and electron waves, respectively, $\tau$ is the
normalized time, $\xi$ is the normalized coordinate and $A$ is the
normalized interaction parameter, which is proportional to the
electron beam current and the length of the active medium.

Solution of the linear equations (\ref{ref_bwopp_01}) and
(\ref{ref_bwopp_02}) predicts unlimited exponential growth of
amplitude of each of the waves. In this case it is necessary to
take into account the nonlinear effects. One of the fundamental
nonlinear effects is the non-isochronism of electrons-oscillators,
expressed in dependence of the frequency of the
electron-oscillator on its energy. Non-isochronism of
electrons-oscillators in the system of interacting waves is
displayed in the shift of nonlinear phase $\varphi_I$ of the
electron wave. Replacing the old variables in the equations
(\ref{ref_bwopp_01}) and (\ref{ref_bwopp_02}) to new ones: $\tau '
= (\tau-\xi)/2 $ and $\xi ' = \xi$ we can rewrite
(\ref{ref_bwopp_02}) as
\begin{equation}\label{ref_bwopp_2}
\frac{\partial
I}{\partial\xi}+j|I|\frac{\partial\varphi_I}{\partial\xi}=-A|F|\exp({j(\varphi_F-\varphi_I))},
\end{equation}
where primes above new variables are omitted.

From the point of view of the non-isochronism effect the
expression for $\partial\varphi_I/\partial\xi $ can be obtained
from the following reasons. The equation for the complete phase
$\varphi_I$ of electron wave must contain the item
$({\omega_e/v_e})\xi$, defining the shift of the frequency of
radiation due to the Doppler effect (here $\omega_e$ is the
frequency of the electron wave and $v_e$ is the longitudinal
velocity of its propagation). So, when we take into account the
dependence of $\omega_e$ upon the energy of the particles ${\cal
W}$, the item proportional to energy ${\cal W}$ should be added at
first approximation into the expression for a slow phase:
\begin{equation}\label{ref_bwopp_3}
 \omega_e=\omega_{e0}+\left({\frac{\partial\omega_e}{\partial{\cal W}}}\right)_0{\cal W}+\dots
\end{equation}

Supposing, that non-perturbed frequency $\omega _ {e0}$ is taken
into account in the "fast" phase, and the energy $ {\cal W} $ of
electrons (in case when their trajectories are identical) is
defined by the energy of the wave by the expression ${\cal W} =
\alpha|I |^2$, where $\alpha$ is the coefficient of
proportionality, we shall discover, that
\begin{equation}\label{ref_bwopp_4}
 \frac{\partial\varphi_I}{\partial\xi}=\frac{\partial\varphi'_I}{\partial\xi}+
 \left(\frac{\partial\omega_e}{\partial{\cal W}}\right)_0{\cal W}\frac{\alpha|I|^2}{v_e}=
 \frac{\partial\varphi'_I}{\partial\xi}+\mu|I|^2,
\end{equation}
where the derivative $\partial\varphi'_I/\partial\xi$ describes
the phase shift of the wave only due to the interaction between
electrons and electromagnetic wave $F$; $\mu=({\alpha}/{v_e})\left
({{\partial\omega_e}/{\partial{\cal W}}}\right)_0 \! \! {\cal W}$
is the parameter characterizing the change of the electron
velocity agreeing with the change of the electron energy during
the interaction. We shall now define the amplitude of an electron
wave as $I=|I|\exp({j\varphi'_I})$ and deal with the new variable
$F'=F/\sqrt{\mu}$, $I'=I/\sqrt{\mu}$, so the equation
(\ref{ref_bwopp_2}) will be written as follows
\begin{equation}\label{ref_bwopp_5}
 \frac{\partial I}{\partial\xi}+j|I|^2I=-AF,
\end{equation}
where the primes above new variable are omitted.

 Thus, the equation (\ref{ref_bwopp_01}) for the backward electromagnetic wave
will not change.

The equations~(\ref{ref_bwopp_01}) and (\ref{ref_bwopp_5})
describe the nonlinear oscillations in the microwave spatially
extended system of interacting backward waves with cubic phase
nonlinearity. Such electron--wave system, as wtitten above, is a
simple model of microwave oscillator with backward wave (BWO \cite
{Soohoo:1971_MicrowaveElectronicsBook, book:1983_microwave} and
gyrotron BWO \cite {book:1983_microwave, Felch:1999_SmallAuthor,
Nusinovich:2001}). This system is suitable for the analysis of
chaotic oscillations, including chaotic synchronization, in
microwave tubes (see, for example, BWO
\cite{Bezruchko:1979_BWOChaos, Ginzburg:1979_engl,
Levush:1992_BWO} and gyrotron backward wave generator
\cite{Nusinovich:2001, Felch:1999_SmallAuthor}).

In an autonomous regime of BWO it is necessary to supplement the
equations~(\ref{ref_bwopp_01}) and (\ref{ref_bwopp_5}) with the
standard boundary conditions $F(\xi=1, \tau)=0$ and
$I(\xi=0,\tau)=0$ describing the lack of electromagnetic and
electron waves on the boundaries of the autonomous system. In the
autonomous microwave oscillator with backward wave at
$\pi/2<A<1.83$ the single-frequency oscillations with stationary
space distribution of electromagnetic field $F(\xi)$ and current
$I(\xi)$ are observed. At $A>1.83$ the multifrequency periodic
oscillations (regime of a periodic self-modulation of an output
field \cite{Ginzburg:1979_engl, Levush:1992_BWO, Nusinovich:2001})
take place. At $A>2.05 $ the time series of the output field
$F(\tau, \xi=0)$ look like pulses with fine oscillations between
them, as a result of excitation of the complex irregular space
distribution $F(\xi) $ and $I(\xi)$. The reason of the latter is
the fast shift of the nonlinear phase of the electron wave along
the space coordinate of the distributed system. At last, at
$A>4.5$ chaotic wide-band oscillations are observed.

In present paper we consider the system of two unidirectionally
coupled, spatially extended electron-wave systems in the chaotic
regime. The system is described by the following equations
\begin{equation}\label{ref_bwopp_01_TWO}
  \frac{\partial F_{1,2}}{\partial \tau}-\frac{\partial F_{1,2}}{\partial \xi}=-{A_{1,2}}I_{1,2},
\end{equation}
\begin{equation}\label{ref_bwopp_5_TWO}
 \frac{\partial I_{1,2}}{\partial \xi}+j|I_{1,2}|^2I_{1,2}=-A_{1,2}F_{1,2},
\end{equation}
where index ``$1$'' corresponds to the drive system and index
``$2$'' - to the response one.

Unidirectional connection between microwave oscillators is brought
into the system as a non-stationary boundary condition for the
slowly varying amplitude of the electromagnetic field $F_2$ of the
response system thus the boundary condition for the first
conducting fissile medium remains constant
\begin{equation}\label{ref_bwopp_boundary_properties_TWO1}
 I_1(\xi=0,\tau)=0,\quad F_1(\xi=1,\tau)=0,
\end{equation}
\begin{equation}\label{ref_bwopp_boundary_properties_TWO2}
 I_2(\xi=0,\tau)=0,\quad F_2(\xi=1,\tau)= \rho F_1(\xi=0,\tau).
\end{equation}
Hence $\rho = R\exp[j\theta]$ is the complex coupling coefficient
($R$ is the coupling strength and $\theta$ is the phase of
coupling coefficient).

For drive and response systems the values of control parameters
are taken as $A_1=4.2 $ and $A_2=4.9$, that corresponds to the
regimes of chaotic oscillations in the active electron--wave
systems with backward waves. The highest Lyapunov exponents are
$\lambda_{A1}=0.224\pm0.010$ and $\lambda_{A2}=0.617 \pm 0.006$
for the first and second control parameter values, respectively.
The value of the phase of coupling coefficient $\theta$ does not
influence the processes in the coupled systems and is fixed
$\theta=\pi$. In present paper the establishment of the chaotic
synchronization regime in two spatially extended electron--wave
systems (BWO) is analyzed at the variation of coupling strength
$R$.

\section{Chaotic synchronization of spatially extended electron--wave systems}
\label{Sct:Synchro}

Let us consider the behavior of the coupled systems with the
increase of amplitude of coupling coefficient $R$ and the fixed
values of control parameters (the values mentioned above are
taken). We shall investigate the oscillations $F_{r1,2}(\tau)=
\textrm{Re}\,\left[{F_{1,2}(\xi=0,\tau)}\right]$ at outputs
$\xi=0$ of both coupled systems.

\begin{figure}
\centerline{\scalebox{0.45}{\includegraphics{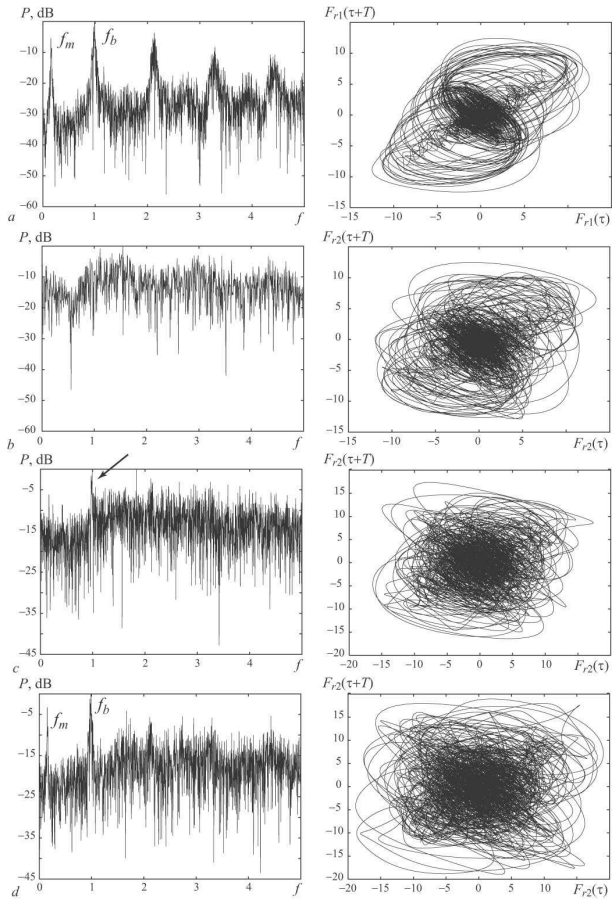}}}
\caption{Power spectra and  phase pictures of output
electromagnetic field oscillations. (a) Drive oscillator with the
value of dimensionless parameter $A=4.2$. (b) Response oscillator
without external signal ($A=4.9$ and $R=0$). (c,d) Response
oscillator with the external signal. The value of coupling
parameter $R$ is (c) $R=0.2$ and (d) $R=0.5$
 \label{fgr:spectrums_R}}
\end{figure}

Figures~\ref{fgr:spectrums_R}{a} and~\ref{fgr:spectrums_R}{b} show
the power spectra and phase pictures (reconstructed by
delay--coordinate embedding method \cite{Packard:1980_Delay,
Takens:1981}) of oscillations at the outputs of drive and response
systems in autonomous regimes (i.e. at $R=0$).

The output field dynamics $F_{r1}$ of the drive oscillator (which
acts upon the input of the interaction space $\xi=1$ of response
oscillator) is chaotic (see Fig.\,\ref{fgr:spectrums_R}{a}). In
the power spectrum the high noise background at the level
$-30\,$dB is observed.
There are two main spectral components with frequencies $f_b$ and
$f_m$ in this spectrum.
High frequency $f_b$ is close to frequency $\hat\omega/2\pi$  of
synchronism of the linear noninteracting electron and backward
electromagnetic waves.
At the decrease of dimensionless parameter of the electron beam
$A_1$ the regime of periodic oscillations with the frequency close
to $f_b$ is established.
The second frequency which is marked as $f_m$ and is settling down
in a low-frequency part of a power spectrum, determines the
low-frequency modulation of complex amplitude $|F_1(\tau)|$ of the
electromagnetic backward wave.

With the growth of control parameter $A$ the oscillations in the
electron--wave system complicate, that is obviously seen from the
analysis of the power spectrum and the chaotic attractor in
Fig.\,\ref{fgr:spectrums_R}({b}), of the second system at $A=4.9$
and $R=0$ (autonomous oscillations).
It is shown in Fig.\,\ref{fgr:spectrums_R}({b}), that the noise
background of the power spectrum rises up to a level
$-(5\div10)$\,dB. With such noise background it is already
impossible to select the main spectral components.

Let us consider the behavior of the response system under the
influence of the drive system signal, at the increase of coupling
coefficient $R$. The growth of the coupling strength parameter $R$
means that the power of external signal acting on the response
system, increases as $R^2$ .

\begin{figure}
\centerline{\scalebox{0.35}{\includegraphics{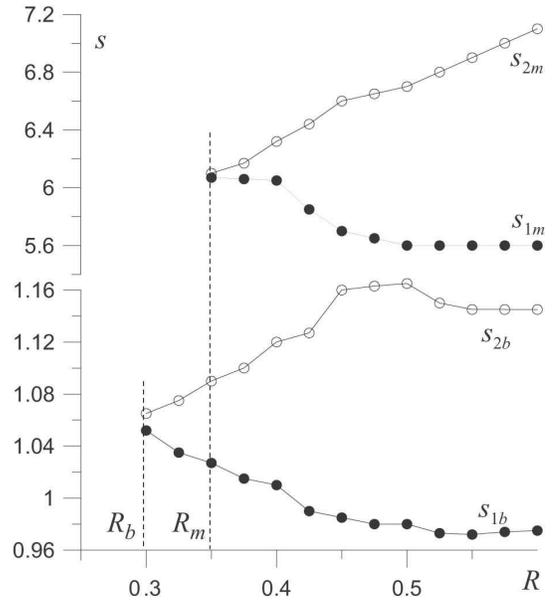}}}
\caption{The dependence of the synchronized time scale range
$[s_1;s_2]$ on the coupling strength $R$ for base time scales
$s_b$ (bottom) and $s_m$ (top)
 \label{fgr:ScaleDifference}}
\end{figure}

With the growth of a coupling parameter $R$ the power spectrum of
generation of the response system changes that is illustrated by
Fig.\,\ref{fgr:spectrums_R}({c,\,d}).
At the small $R<0.3$ [Fig.\,\ref{fgr:spectrums_R}(c)] upon the
noise base of the response system spectrum  there is a weakly
delineated frequency (marked by an arrow), corresponding to the
base frequency $f_b $ of the driving system spectrum [see
Fig.\,\ref{fgr:spectrums_R}(a)]. At the large values $R>0.3$ the
energy of the spectral component corresponding to the base
frequency (marked by arrow in Fig.\,\ref{fgr:spectrums_R}(d);
$R=0.5$) increases.

It is necessary to mark, that both systems have attractors with
ill-defined phase. For these chaotic signals it is impossible to
introduce the instantaneous phase $\phi(t)$ of chaotic signal
correctly. It is clear that in such case the traditional methods
of the phase synchronization detecting (see Ref.
\cite{Rosenblum:1996_PhaseSynchro, Anishchenko:2004_ChaosSynchro,
Anishchenko:2002_SynchroEng, Pikovsky:2000_SynchroReview}) fail.
On the contrary, the approach based on introduction of
instantaneous phase $\phi_s(t)$ by the wavelet transform
\cite{Koronovskii:2004_JETPLettersEngl, Hramov:2004_Chaos} can
easily help to detect the time scale synchronization between
chaotic oscillators.

The dependence of synchronized time scale range $[s_1;s_2]$ on
coupling parameter $R$ is shown in Fig.~\ref{fgr:ScaleDifference}.
The appearance of the synchronized time scale range corresponds to
the TSS regime.

The peculiarity of dynamics of the system of two unidirectional
coupled spatially extended systems is that the TSS of chaotic
oscillations of each of subsystems at a large value of coupling
parameter $R$ is observed for two ranges near to base time scales
$s_b=1/f_b $ and $s_m=1/s_m $, respectively. In
Fig.\,\ref{fgr:ScaleDifference} the ranges of phase locked scales
are shown for the areas near to the base time scale $s_b$ (lower
figure), and the time scale $s_m $ (upper figure).

From Fig.\,\ref{fgr:ScaleDifference} it can be seen, that for
coupling coefficient values $R<0.29$ the establishment of a
chaotic synchronization regime is not observed. The latter means,
that there are no time scales $s $ which would satisfy the
conditions of TSS (\ref{eq:PhaseLocking}) and
(\ref{eq:SynchroEnergy}).

\begin{figure}
\centerline{\scalebox{0.39}{\includegraphics{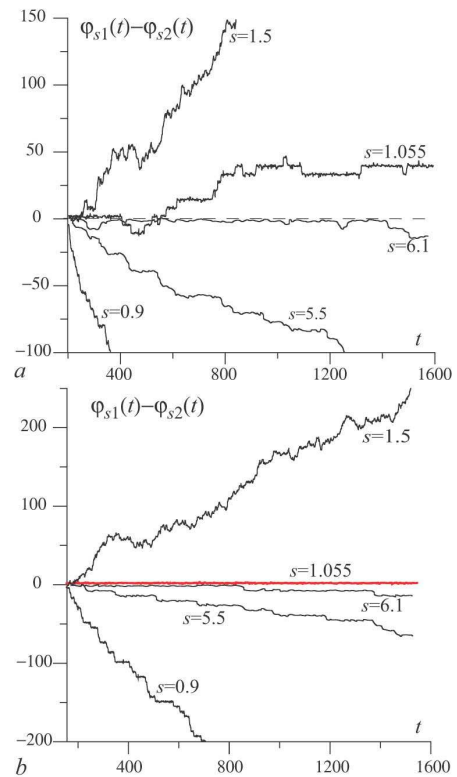}}}
\caption{The phase difference $\phi_{s1}(t) - \phi_{s2}(t)$ for
two unidirectionally coupled spatially extended electron--wave
systems. (a) $R=0.2$. There is no synchronization between systems.
(b) $R=0.3$. The time scales $s=1.055$ (red line) of both systems
are correlated with each other and the time scale synchronization
regime can be observed
 \label{fgr:WaveletPhases}}
\end{figure}

Figure~\ref{fgr:WaveletPhases} illustrates the dynamics of phase
difference $\phi_{s1}(t)-\phi_{s2}(t)$ for different time scales
for $R=0.2$. It is clear, that there is no time scale range
$[s_1;s_2]$ for which the phase difference $\phi_{s1}(t) -
\phi_{s2}(t)$ ($s\in[s_1;s_2]$) is bounded.

For the coupling parameter  $R$ exceeding the critical value
$R_c\simeq 0.29$ we observe the time scale range $\Delta
s=s_2-s_1$ for which the requirement of capture of phases
(\ref{eq:PhaseLocking}) is satisfied and the energy relating on
this range (\ref{eq:SynchroEnergy}) is nonzero. The latter means,
that in the coupled active media the TSS regime is established
\cite{Koronovskii:2004_JETPLettersEngl, Hramov:2004_Chaos}.

The behavior of the phase difference $\phi_{s1}(t) - \phi_{s2}(t)$
for this case is presented in Fig.\,\ref{fgr:WaveletPhases}(b).
So, we can say, that the time scales $s=1.055$ (and close to them)
of chaotic oscillations of electromagnetic field amplitudes on the
output of coupled systems are synchronized with each  other. At
the same time the other time scales (e.g., $s=1.5$) remain
uncorrelated. For such time scales the phase locking is not
observed [see Fig.~\ref{fgr:WaveletPhases}(b)].

Let us note, that with the growth of coupling parameter the TSS
appears, at first, near the base scale $s_b$ at $R=R_b$ (it is
marked in Fig.\,\ref{fgr:ScaleDifference}). With the further
increase of coupling coefficient, at $R=R_m> R_b $ the phase
locking of time scales occurs in the area of a time scale $s_m$,
corresponding to modulating frequency $f_m$ of amplitude of the
output field of the response BWO.

\begin{figure}
\centerline{\scalebox{0.7}{\includegraphics{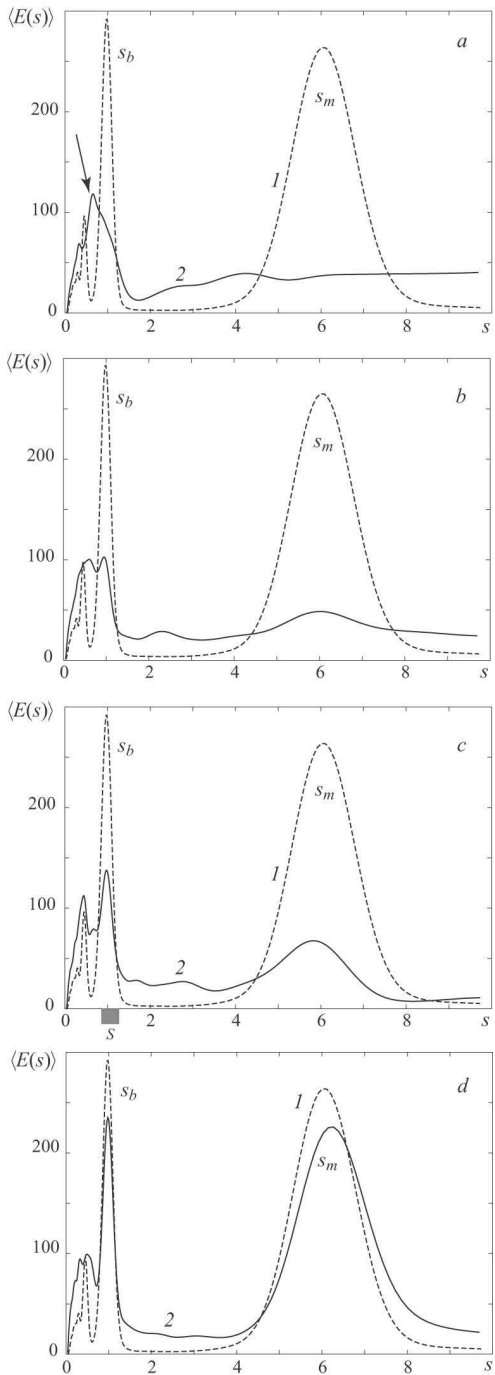}}}
\caption{The energy distribution in wavelet spectra $\langle
E(s)\rangle$ for output field oscillations of drive (stroke line)
and response (solid line) spatially extended systems. (a) $R=0$.
There are autonomous oscillations in the systems. (b) $R=0.2$.
There is no time scale synchronization between systems. (c)
$R=0.3$. There is time scale synchronization near the base time
scale $s_b$ (Synchronized scale region is marked by gray box $S$).
(d) $R=0.5$. Time scale synchronization near both base time scale
$s_b$ and modulation time scale $s_m$ is observed.
 \label{fgr:WaveletSpectra}}
\end{figure}

Let us consider at first, why the range $\Delta s$ of synchronized
time scales (i.e. time scales, for which the conditions of
synchronization (\ref{eq:PhaseLocking}) and
(\ref{eq:SynchroEnergy}) are fulfilled) appears near the time
scale $s_b$ corresponding to the base frequency $f_b=1/s_b $.
For this purpose we shall study wavelet power spectra $ \langle {E
_ {1,2} (s)} \rangle $, defined from the equation
(\ref{eq:IntEnergy}), for drive and response systems.
Wavelet spectra of the output field oscillations of drive (a
stroke line~{\sl1}) and response (a solid line~{\sl2}) BWO are
presented in Fig.\,\ref{fgr:WaveletSpectra} for different values
of coupling parameter $R$.
Let us mark, that the wavelet power spectrum
$\langle{E_1(s)}\rangle$ of the drive system does not vary (since
the value of controlling parameter $A_1 $ is fixed and the system
is autonomous) and is presented in all the figures, for different
values of $R$, as a matter of convenience comparison with the
wavelet power spectra $\langle{E_2}\rangle$ of the response
system.

In Fig.\,\ref{fgr:WaveletSpectra}({a}) wavelet power spectra of
autonomous oscillations are shown ($R=0$).
One can see that the oscillations in each of the spatially
extended systems have essentially different wavelet power spectra.
So, in wavelet spectrum $\langle{E_1(s)}\rangle$ of the drive
system the base time scales $s_b $ (corresponding to the high
frequency $f_b$) and $s_m$ (corresponding to modulation
oscillations with the frequency $f_m=1/s_m$) stand out against the
background. In wavelet spectrum of an autonomous response system
the time scales corresponding to high-frequency spectral component
(they are marked by an arrow in Fig.\,\ref{fgr:WaveletSpectra}(a))
also precisely stand out against background. However, the
arrangement of a peak of $s_{b}$ in wavelet power spectrum
$\langle {E_2(s)}\rangle $ does not correspond to a base time
scale $s_b$ of the drive system.
As to large time scales (corresponding to low modulation
frequencies), their wavelet power spectra are essentially
different.
The wavelet spectrum of the response system in the area of the
time scale $s_m$ is continuous and homogeneous, i.e. the
modulation oscillations are characterized by continuous noise-like
spectrum.
For small values of coupling coefficient $R$, the shape of wavelet
power spectrum changes [see Fig.\,\ref{fgr:WaveletSpectra}({b})).
However, time scale synchronization is not observed, which is
apparent from Fig.\,\ref{fgr:WaveletPhases}(a).

With the increase of $R$ the conditions of TSS
(\ref{eq:PhaseLocking}) and (\ref{eq:SynchroEnergy}) are satisfied
only by the time scales $s$ whose energy in power spectra of the
drive and response systems is rather large. Really, at the values
of coupling coefficient $R\approx0.3 $ as it was already
considered above, the conditions of TSS are fulfilled only for
time scales near the basic scale $s_b=1/f_b\approx1.0 $ (see also
Fig.\,\ref{fgr:ScaleDifference}).
As it follows from Fig.\,\ref{fgr:WaveletSpectra}(c), in the range
of time scales marked in Fig.\,\ref{fgr:WaveletSpectra}(c) by grey
color and symbol $S$, the energy of wavelet power spectrum is
maximum.

From Fig.\,\ref{fgr:WaveletSpectra}(c) one can also see that the
energy of wavelet spectrum, being fallen on the time scale $s_m $
of the response system simultaneously increases. However, at
$R\approx0.3$ its energy is not large enough, and TSS of field
modulation oscillations is not observed.

At large coupling coefficients $R> 0.35 $ the shape of wavelet
power spectra of output field oscillations of response and drive
systems become close to each other (see Fig.\,\ref
{fgr:WaveletSpectra}(d)).
Base time scales $s_m$ and $s_b$ are presented in the power
spectra with approximately equal intensity. The analysis of time
scales dynamics shows, that the conditions of synchronization
(\ref{eq:PhaseLocking}) and (\ref{eq:SynchroEnergy}) are fulfilled
for both most intensive time scales in wavelet power spectra and
two ranges of time scales, in which phase locking occurs [see
Fig.\,\ref{fgr:ScaleDifference} for $R> 0.35$], are accordingly
selected.

Thus, the TSS of two coupled chaotic spatially extended systems is
observed first of all for those time scales $s$ whose energy in
the chaotic power spectrum is great.

With the increase of coupling parameter the range of synchronized
time scales $[s_1, s_2]$, for which the TSS conditions
(\ref{eq:PhaseLocking}) and (\ref {eq:SynchroEnergy}) are
satisfied, is expanded. Time scales which are close to the most
intensive scales in wavelet power spectrum and whose energy is
large enough, start to be involved in synchronous dynamics.

The latter is well illustrated by Fig.\,\ref{fgr:ScaleDifference}
from which it can be seen, that for both base time scales $s_b$
and $s_m$ extension of the range of phase locked scales $\Delta s$
with the increase of $R$ is observed.
At large values of coupling coefficient $R$ the range of
synchronous scales $\Delta s_b=s_{2b}-s_{1b}$ in the area of the
base scale $s_b$ ceases to be expanded and, moreover, is narrowed
down at $R>0.5$.
At the same time in the area of scales $s_m$, describing
modulation of amplitude of the output field of BWO, the range of
synchronous scales $\Delta s_m=s_{2m}-s_{1m}$ increases linearly
with the growth of $R$, at the expense of growth of top bound of
area of synchronized scales $s_{2m}$.
At last we must note that with growth of $R$, the energy of the
wavelet power spectrum of a response system is redistributed in
such a manner that the increase of energy of larger scales $s>s_m
$ is observed.
Then the most intensive time scales in wavelet spectrum appear
locally, near the top bound $s_{2m}$ of synchronized time scales,
therefore these time scales are first to be synchronized with the
increase of coupling coefficient $R$.

\section{Measure of synchronization of spatially extended electron--wave system}
\label{Sct:Measure}

Introducing continuous set of time scales $s$ and the
instantaneous phases associated with them, as well as separation
of the range of synchronous time scales $\Delta s=s_2-s_1$ allows
us to inject the quantitative performance of a measure of chaotic
synchronization of coupled systems
\cite{Koronovskii:2004_JETPLettersEngl, Hramov:2004_Chaos}. This
measure $\gamma$ can be defined as a part of wavelet spectrum
energy being fallen on the synchronized time scales
\begin{equation}
\gamma=\frac{1}{E_{2}}\int\limits_{s_m}^{s_b}\langle
E_{2}(s)\rangle\,ds, \label{eq:Gamma}
\end{equation}
where $[s_m;s_b]$ is the range of time scales for which the
condition~(\ref{eq:PhaseLocking}) is fulfilled and $E_{2}$ is the
full energy of the wavelet spectrum
\begin{equation}
E_{2}=\int\limits_{0}^{+\infty}\langle E_{2}(s)\rangle\,ds
\end{equation}
of the response system.

The energy distribution over different time scales is a very
useful characteristics for complex system behavior description. It
is widely used in different applications (see e.g.,
\cite{Schwarz:1998_WaveletsAstro, WaveletsInPhysics:2004}).

This measure $\gamma$ is equal to zero for the nonsynchronized
oscillations. $\gamma\not=0$ means that in the coupled systems the
conditions of TSS (\ref{eq:PhaseLocking}) and
(\ref{eq:SynchroEnergy}) are implemented. The value $\gamma=1$
shows, that oscillations in each of subsystems are identical or
shifted upon some lag of time $T_\tau$. This is a complete or lag
chaotic synchronization \cite{Pikovsky:2002_SynhroBook}. Increase
of $\gamma$ from $0$ up to $1$ testifies to the increase of a part
of the wavelet spectrum energy, being fallen on synchronous time
scales $s$. Actually, the measure $\gamma$ characterizes how close
to each other the chaotic oscillations are in both spatially
extended systems.

In our case when the synchronous behaviour is observed for two
base scales $s_m$ and $s_b$, it is possible to introduce measures
of synchronization $\gamma_b$ and $\gamma_m$ according to each of
time scales $s_b$ and $s_m$, and also the integral performance of
the TSS measure of the systems: $\gamma=\gamma_{b} + \gamma_{m}$.
Dependence of the measure $\gamma$ on the coupling coefficient $R$
characterizes degree of chaotic synchronization of oscillations in
the electron--wave systems with a backward wave.

\begin{figure}
\centerline{\scalebox{0.4}{\includegraphics{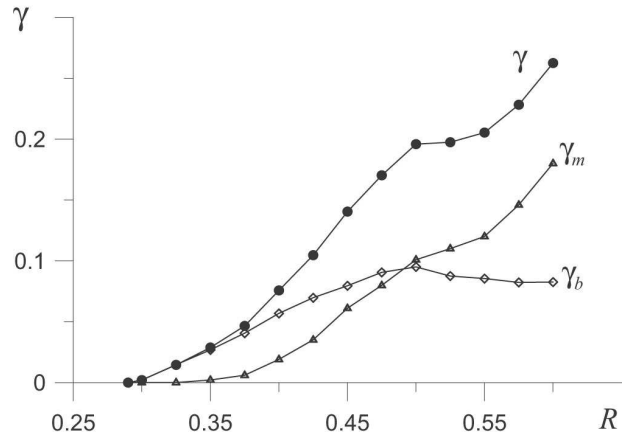}}}

\caption{The dependence of the synchronization measures
$\gamma_b$, $\gamma_m$ and $\gamma$ of the response system on the
coupling strength $R$
 \label{fgr:Gamma}}
\end{figure}

In Fig.\,\ref{fgr:Gamma} the dependence of the synchronization
measures $\gamma_b$, $\gamma_m$ and $\gamma$ for the response
systems upon the coupling parameter $R$ is shown. It is obvious,
that at small $R<0.4\div0.5 $ the main part of energy of the
wavelet power spectrum, being fallen at synchronized time scales,
corresponds to the base scale $s_b$ ($\gamma_b>\gamma_m$).

Thus, in the coupled systems there is the capture of phases of
time scales $s\sim s_b$ close to the time scale corresponding to
the frequency of synchronism of electronic and electromagnetic
waves. With further growth of $R$ the capture of frequencies close
to the modulating frequency $f_m$ of the output signal of the
driving BWO takes place. With the increase of $R$ the range of the
synchronized scales describing low-frequency chaotic modulation of
the output signal also increases. Hence, $\gamma_m$ (the
normalized part of energy of synchronized scales corresponding to
low-frequency modulation oscillations)  grows with the increase of
$R$. At $R> 0.5$ part of energy of synchronous low-frequency
modulation time scales $\gamma_m$ becomes more than the value
$\gamma_b$, describing part of energy of synchronous
high-frequency oscillations.

The integral part of energy $\gamma$, being fallen on synchronous
time scales, monotonically increases with the growth of value of
coupling coefficient $R$ [see Fig.\,\ref {fgr:Gamma}]. However,
even at large values of $R$, when the response active medium is
hardly influenced by the signal of the driving system
($R^2\sim0.3\div0.4 $), the part of the energy, being fallen upon
synchronous time scales, does not exceed value $\gamma\sim0.3 $.

Thus, at chaotic synchronization of two unidirectional coupled BWO
at the increase of coupling coefficient, we can observe TSS near
the most intensive time scale $s_b$ corresponding to frequency
$f_b$ of synchronism of electronic and electromagnetic waves. With
the growth of $R$ there is an increase of energy of synchronized
scales in the area of the base scale $s_b$. At the same time,  a
range of synchronized time scales $\Delta s_m$, corresponding to
low-frequency modulation of oscillations of field $F$, appears.

With the further increase of $R$ the growth of wavelet power
spectrum energy, being fallen on synchronous time scales, is due
only to the increase of the range of synchronized time scales
$\Delta s_m$. The given behaviour of two unidirectional coupled
systems is quantitatively described by the measure of chaotic
synchronization $\gamma=\gamma(R)$.

\section{Space--time dynamics of response chaotic electron--wave system}
 \label{Sct:SpaceDynamics}

Let us consider the physical processes in the non--autonomous
(response) active medium when the chaotic synchronization appears.
First of all, we are interested in space dynamics of response
active medium.

The external chaotic signal generated by the drive distributed
system influences on the point with coordinate $\xi=L$ of the
response active medium and then spreads towards electron wave to
the output $\xi=0$ border. The dimensionless length of the active
media denoted as $L$ is equal to $1$. The synchronization of the
whole active electron--wave system discussed above means the time
scale synchronization of external signal $F_{ext1}({\tau})$ and
output signal $F_{out2}(\tau)=F_2(\tau,\xi=0)$ of response media
takes place. To consider the space--time dynamics of response
system we have analyzed the presence of time scale synchronization
between the external signal and the time series $F_2(\tau,\xi)$
obtained from different points $0<\xi<L$ of interaction space.

Let us consider the synchronization on time scales which are close
to the most intensive scale $s_b$ of external chaotic signal
wavelet spectrum. The synchronization of these time scales appears
when the value of coupling strength $R$ is small enough and when
the energy of external signal $P_{ext}=R^2|F_{ext1}|^2$ is also
small (see Section~\ref{Sct:Synchro} for detail). The range
$\Delta s_b$ of synchronized time scales may be considered as a
function of interaction space coordinate of active medium. This
function ${\Delta s_b=\Delta s_b(\xi)}$ as well as the
synchronization measure ${\gamma=\gamma(\xi)}$ (\ref{eq:Gamma})
are the characteristics describing the space dynamics of chaotic
synchronization appearance when the external chaotic signal acts
on one of the border of the distributed system.

\begin{figure}
\centerline{\scalebox{0.5}{\includegraphics{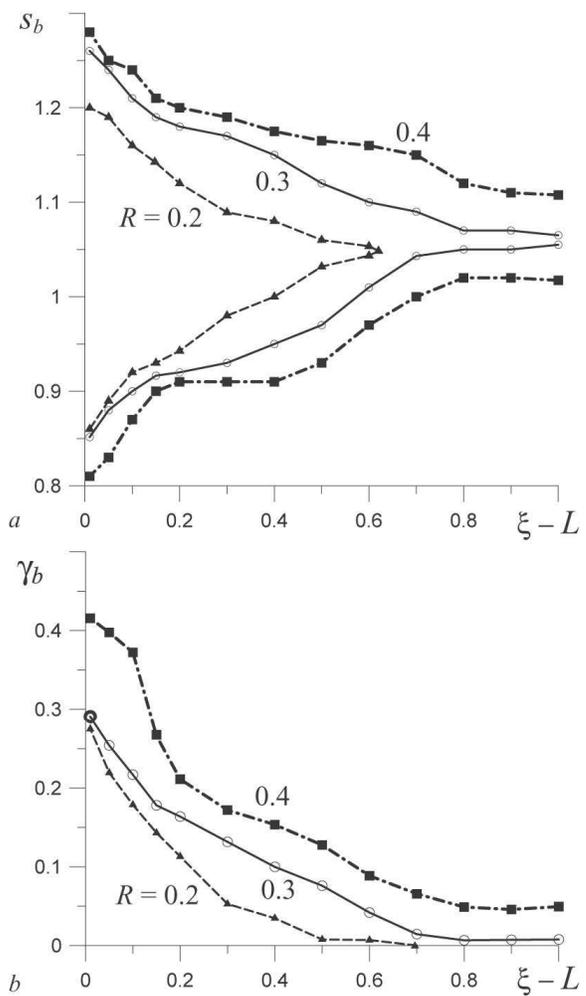}}}
\caption{(a) The areas of synchronized time scales on the plane
``coordinate $(L-\xi)$ --- time scales $s$''. The coordinate
${(L-\xi)}$ is counted from the end of the response system. (b)
The dependence of synchronization measure $\gamma_b$ on the
coordinate of interaction space $(L-\xi)$. The black squares
($\blacksquare$) correspond to the value of coupling parameter
$R=0.4$, circles ($\circ$) --- $R=0.3$ and black triangles
($\blacktriangle$) --- $R=0.2$ \label{fgr:DsG_xi}}
\end{figure}

Fig.~\ref{fgr:DsG_xi}({a}) shows areas of synchronized time scales
on the plane ``coordinate $(L-\xi)$ --- time scales $s$'' for
three different values of coupling strength $R$. In
Fig.~\ref{fgr:DsG_xi}(a) one can see that the width of the
synchronized time scales area is maximum in the points of
interaction space which are close to the border of system on which
the chaotic signal of the drive active medium acts (see the border
conditions (\ref{ref_bwopp_boundary_properties_TWO2}) for the
field $F_2$ of the response system). When the coordinate $\xi$
decreases, the area of synchronized time scales $\Delta s_b(\xi)$
converges gradually. The measure of synchronization
$\gamma_b(\xi)$ behaves along the interaction space in the similar
way [see Fig.~\ref{fgr:DsG_xi}(b)]. With the decreasing of
$\xi$--coordinate the part of wavelet spectra energy being fallen
on the synchronized time scales $\Delta s_b(\xi)$ also decreases.

When the coupling strength $R$ is rather small ($R<0.29$) and the
synchronization of two coupled active media is not observed [see
Sections~\ref{Sct:Synchro} and \ref{Sct:Measure}], the width of
the synchronized time scales area becomes equal to zero at the
point of interaction space $\xi_s<L$. It means that the intricate
space dynamics of phase locking takes place in the asynchronous
regime. The dependence of the width of the synchronized time
scales area on the coordinate $\xi$ (see also
Fig.~\ref{fgr:DsG_xi}(a), case $R=0.2$) is described by
\begin{equation}
 \begin{array}{ll}
   \Delta s\not=0  &
   \text{~при~}\xi>\xi_s, \\
   \Delta s=0  & \text{~при~}\xi<\xi_s. \\
 \end{array}
\label{eq:Ds_xi}
\end{equation}
So, in this case it is possible to say that the interaction space
of the response active medium is separated to two regions.

First of them is characterized by length $L_s=L-\xi_s$ and abuts
on the border $\xi=L$ of the response active system. The chaotic
oscillations $F_2(\tau,\xi)$ ($\xi\in(\xi_s,L)$) in this region
are synchronous (in terms of time scale synchronization) with the
external chaotic signal $F_{ext1}(\tau)$ generated by the drive
system. Therefore, the synchronization measure is not equal to
zero: $\left.\gamma_b\right|_{\xi>\xi_s}\not=0$. This region of
interaction space has naturally been called as ``the region of
synchronous oscillation''. Its length $L_s$ has been called
``length of synchronization''.

In the second region $\xi\in(0,\xi_s)$ the chaotic synchronization
is destroyed and the range of the synchronized time
scales~(\ref{eq:Ds_xi}) as well as the synchronization measure is
equal to zero.

With the increasing of $R$ the synchronization length also
increases. For some value $R=R_b$ the length of synchronization
becomes equal to the length of the interaction space $\L_s\equiv
L$ (see Section~\ref{Sct:Synchro} and
Fig.~\ref{fgr:ScaleDifference}). This means that in all space of
response of electron--wave medium the chaotic synchronization
takes place and, therefore, one can say about synchronization of
considered systems as whole. This regime of chaotic
synchronization has been described above in
sections~\ref{Sct:Synchro} and \ref{Sct:Measure} when the time
scale dynamics of chaotic signal get from output $\xi=0$ of
response system has been considered.

Fig.~\ref{fgr:DsG_xi}(a) corresponding to the case of
synchronization of the whole response active medium shows the
borders of synchronized time scales $s$ along the interaction
space. The value of coupling strength has been selected as
$R=0.3\approx R_b$. There is the range $\Delta s$ of synchronized
time scales for every point $0\leq\xi\leq L$. With coupling
strength increasing the range of synchronized time scales also
increases [see Fig.~\ref{fgr:DsG_xi}(a), $R=0.4$]. This result
agrees well with Sec.~\ref{Sct:Synchro}. The value of wavelet
power spectrum energy being fallen on synchronized time scales is
equal approximately to $30\div40\%$ in the point with coordinate
$\xi=0.99L$ and decreases to $1\div5\%$ at the output border
$\xi=0$ of the response active media.

\begin{figure}
\centerline{\scalebox{0.3}{\includegraphics{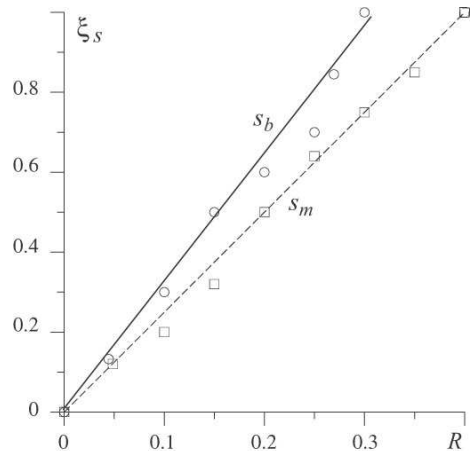}}}
\caption{Length of synchronization of response spatially extended
system as function of coupling strength $R$. The dependence for
base time scale $s_b$ is pointed by circles and approximation
$L_s\sim R$ is plotted by solid line.  The numerical dependence
and approximation for modulating time scale $s_m$ is pointed by
squares and dashed line \label{fgr:Xis_R}}
\end{figure}

In Fig.~\ref{fgr:Xis_R} the dependence of the length of
synchronization $L_s$ on coupling parameter $R$ for the base time
scale $s_b$ is shown (solid line). It is clear, that with the
increasing of coupling strength the length of synchronization
grows monotonically according to the linear law: $L_s(R)\sim R$.
It follows, that with the growth of coupling coefficient $R$ we
have the linear increase of the region where time scale
synchronization near base scale $s_b$ is observed. Such behaviour
of the response system is determined by the features of
interaction of electronic and electromagnetic waves. Destruction
of time scale synchronization is defined by violation of phase
ratio between backward waves  of electron current $I(\tau,\xi)$
and electromagnetic field $F(\tau,\xi)$.

So, at the fixed length of the response system, the phase shift
$\Delta\varphi$ between field and current waves, originating due
to the influence of external signal upon the response system,
reduces on the length of synchronization $L_s$ to destruction of
time scale synchronization on some time scale $s$. The phase shift
is defined as $\Delta\varphi=2\pi|f_b-f_s|\xi_s/v_g$, where
$f_b=1/s_b $, $f_s=1/s$ and $v_g$ is the the group velocity of the
backward electromagnetic wave.

On the other hand, dependence of the phase shift $\Delta\varphi$
upon the amplitude of the external field $R|F_2|$ can be presented
as $\Delta\varphi=\kappa RF_2$, where $\kappa$ is the coefficient
of proportionality depending on length of the response spatially
extended system $A_2$. Comparing the estimated expressions for the
phase shift $\Delta\varphi$ it is found out that the length of
synchronization $L_s$ is proportional to $R$, as the numerical
simulation shows (see Fig.~\ref{fgr:Xis_R}).

It is important to note that curves shown in Fig.~\ref{fgr:DsG_xi}
correspond to the base time scale $s_b$ and time scales close to
it. The similar results have been obtained for the space--time
dynamics of scales which are close to the modulation time scale
$s_m$ [see Fig.~\ref{fgr:Xis_R}, dashed line]. The difference
consists in that in this case the length of synchronized time
scales area (length of synchronization) becomes equal to the
system length for the bigger values of coupling strength than for
the base time scale $s_b$. This also agrees well with the results
discussed above.

\section{Conclusion}
\label{Sct:Conclusion}

In conclusion, in our article we have considered the chaotic
oscillations in two unidirectionally coupled electron--wave active
media. The approach of chaotic synchronization analysis proposed
in our work~\cite{Koronovskii:2004_JETPLettersEngl,
Hramov:2004_Chaos} has been used. This approach is based on the
consideration of continuous set of time scales and chaotic signal
phases associated with them. The application of this method allows
us to investigate the chaotic synchronization of distributed
active media demonstrating the intricate dynamics whereas the
traditional methods of chaotic signal phase introduction fail due
to the complex topology of the attractor. At the same time we
succeed in dividing the chaotic synchronization into
synchronization of low--frequency modulation oscillations of
electromagnetic field and high--frequency generation one. One can
see that it was impossible to do it using traditional approaches
of chaotic synchronization analysis. So, the results described in
our work allow to understand more clearly the common mechanisms of
chaotic synchronization and justify the efficiency of new
TSS-approach to chaotic synchronization analysis.

The main results are the following. With the increase of coupling
strength the chaotic synchronization regime appears and is
characterized by phase relationship between time scales of chaotic
oscillations of drive and response active media. First of all, the
synchronization of time scales corresponding to high--frequency
spectral component appears. The synchronization of time scales
corresponding to low--frequency oscillations takes place with the
larger values of coupling strength $R$. When the chaotic
synchronization of the whole active medium is not observed the
interaction space may be divided into two regions corresponding to
synchronous and asynchronous oscillations, respectively. The
length of synchronous oscillation area $L_s$ grows when the
coupling strength $R$ increases. The appearance of chaotic
synchronization of two unidirectionally coupled active media is
determined by condition of equality of synchronization length and
length of the system.

It should also be noted that the method of investigation of
chaotic synchronization of two coupled distributed systems
discussed above can be easily applied to the analysis of higher
spatial dimensions systems, such as 2D and 3D ones. In this case
one have to analyze time series $x(t,\mathbf{r})$ obtained from
different points $\mathbf{r}$ of space in the manner discussed
above.

\section{Acknowledgments}
\label{Sct:Acknowledgments} We express our appreciation to
Professor Dmirtiy~I.~Trubetskov for valuable discussions. We also
thank Dr. S.V.~Eremina for language support.
This work has been supported by U.S.~Civilian Research \&
Development Foundation for the Independent States of the Former
Soviet Union (CRDF, grant {REC--006}) and Russian Foundation for
Basic Research (grant 05--02--16273). We also thank ``Dynasty''
Foundation.


\end{document}